\documentclass[useAMS,usenatbib,fleqn]{mn2e}

\usepackage[pdftex]{graphicx}
\usepackage[outdir=./PDFFigs/]{epstopdf}
\usepackage[T1]{fontenc}
\providecommand{\adsurl}[1]{\href{#1}{ADS}}

\title[O$_2$ in comet 67P/Churyumov-Gerasimenko]
{A gas-phase primordial origin of O$_2$ in comet 67P/Churyumov-Gerasimenko}
\author[J. M. C. Rawlings, T. G. Wilson and D. A. Williams]
{J. M. C. Rawlings$^1$\thanks{E--mail: jcr@star.ucl.ac.uk} 
T. G. Wilson$^{1}$ and D. A. Williams$^1$\\ 
$^1$Department of Physics and Astronomy, University College London, Gower Street, London, WC1E 6BT.}
\begin{document}

\maketitle

\begin{abstract}

\noindent Recent observations made by the \textit{Rosetta}/ROSINA instrument have 
detected molecular oxygen in the coma of comet 67P/Churyumov-Gerasimenko with 
abundances at the 1-10\% level relative to H$_2$O. 
Previous studies have indicated that the likely origin of the O$_2$ may be surface 
chemistry of primordial (dark cloud) origin, requiring somewhat warmer, denser
and extreme H-atom poor conditions than are usually assumed. 
In this study we propose a primordial 
gas-phase origin for the O$_2$ which is subsequently frozen and effectively hidden
until the ice mantles are sublimated in the comet's coma. Our study  presents 
results from a three-phase astrochemical model that simulates the chemical 
evolution of ices in the primordial dark cloud phase, its gravitational collapse, and 
evolution in the early protosolar nebula. We find that the O$_2$ abundance can be 
produced and is fairly robust to the choice of the free parameters. Good matches
for the O$_2$:H$_2$O ratio and, to a lesser extent, the N$_2$:CO and CO:H$_2$O 
ratios are obtained, but the models 
significantly over-produce N$_2$. We speculate that the low value of N$_2$:O$_2$ that
is observed is a consequence of the specific thermal history of the comet.

\end{abstract}


\section{Introduction}
\label{sec:intro}

Comets are believed to be among the most primitive objects in the
Solar System and potentially carry within them the chemical signatures of 
molecular clouds from which the nascent protosolar nebula (PSN) formed. However, cometary material 
and ices may be created by and subject to a wide variety of chemical processing 
mechanisms including; chemistry in the  
molecular cloud as well as the PSN, solid-state chemistry in or on the ices, heating
by collisions as well as internal radiogenic heating, surface radiation processing by 
impacting high energy particles (radiolysis), condensation/sublimation and the effects 
of Solar irradiation (photolysis).
 
Comet 67P/Churyumov-Gerasimenko (hereafter referred to as comet 67P/C-G) is a Jupiter-family 
comet, which is thought to have originated from the trans-Neptunian Kuiper belt, but has 
been perturbed by
a close encounter with Jupiter (in 1959) into an inner Solar-System orbit \citep{Maquet15}. 
These comets do not form a homogenous class and significant 
chemical variations exist; for example the D/H ratio in 67P/C-G is $> 3\times$ larger than in other 
Jupiter-family comets, such as Hartley 2, \citet{Bal15}, and the orbits/dynamical histories
may be quite different (e.g. 67P/C-G has had just 10 apparitions in its present orbit, whilst 
the short-period comet P/Halley has had $>$2000 apparitions, \citealp{Hughes85}).

Comet 67P/C-G has merited detailed attention in the wake of the \textit{Rosetta} mission. 
This yielded many interesting and unexpected results, not least the realisation that 
comets of this type cannot be represented by `dirty snowballs' as previously thought, but are
essentially `dry' with ices and volatiles amounting to $\leq$15\% of the cometary mass, embedded 
within a rocky matrix \citep{Fulle16}.
The outgassing from the comets is believed to be characterised by the sublimation of 
successive layers of ancient icy material (in onion-skin fashion) in the vicinity of 
perihelion on each orbit \citep{GLetal15}. 

One of the many intriguing results from the \textit{Rosetta} mission has been the discovery of a
larger-than-expected abundance of molecular oxygen (O$_2$) in the cometary coma. 
Cometary gases are typically primarily composed of H$_2$O, CO and CO$_2$ with small 
traces of other atomic and molecular species \citep{BMetal04}. However, the ROSINA-DFMS device 
(Rosetta Orbiter Spectrometer for Ion and Neutral Analysis - Double Focusing Mass Spectrometer) 
determined the presence of O$_2$ at the 1-10\% level relative to H$_2$O, with a 
mean value of $\sim3.80\%$; 
significantly larger than the gas-phase values determined for interstellar dark 
clouds and protostellar sources (e.g. O$_2$:H$_2$O$<6\times 10^{-9}$ in NGC1333-IRAS4A, \citealp{Yildiz13}). 
This is well below the observational upper limits of 15\% and 39\% for O$_2$:H$_2$O in 
the interstellar ices associated with low and high star-forming regions, respectively 
\citep{BGW15}, although it is interesting to note that the O$_2$:CH$_3$OH ratio in the warm
(sublimated) gas of IRAS 16293-2422 is lower than that in 67P/C-G \citep{TQ18}.
Moreover, the O$_2$ appears to be distributed isotropically and the O$_2$:H$_2$O ratio is
not obviously correlated to the heliocentric distance of the comet. Nor is the O$_2$
well-correlated with N$_2$ and CO, molecules of similar volatility/binding 
energy \citep{Bieler15}.

This suggests that the O$_2$ permeates the cometary ice to some depth 
and is not merely confined to the surface layers. It has been shown that erosion 
of 67P/C-G causes loses of metres to tens of metres of material per apparition 
\citep{Keller15} so that in its 10 orbits, 67P/C-G will have lost surface material
up to depths of hundreds of metres.

Studies based on the structure, morphology and composition of 67P/C-G
indicate that is essentially a `primordial rubble pile' formed by the
agglomeration of material in the protosolar nebula 
that remained after the formation of trans-Neptunian objects (TNOs)
as opposed to `collisional rubble piles' formed as a result of the 
disruption of larger parent bodies \citep{WAL04,Detal16}. 
The comet is thus believed to have 
grown slowly at very low temperatures, from aged material in the inner 
Solar System, and avoided thermal processing by the decay of short-lived 
radionuclides (e.g. $^{26}$Al) or collisional processing 
- that would otherwise lead to a loss of the volatile species, 
such as CO, CO$_2$ and O$_2$. This agglomeration could have been 
hierarchical \citep{Detal16,Bent16}, or even 
just via a gentle assembly of mm-sized `pebbles' and
ices \citep{Blum17}.
Moreover, the bi-lobe structure of 67P/C-G probably resulted from a very 
low velocity (<1ms$^{-1}$) collision \citep{Mass15,Blum17} and that collision 
possibly occurred within the last 1Gy \citep{JBTM16}.

As the significance of the delivery of cometary H$_2$O to terrestrial oceans continues to be debated 
\citep{Altwegg15,Willacy15}, 
we need to understand the origin and evolution of the cometary ices. It is possible 
that ratios of simple molecular species, such as O$_2$:H$_2$O, can provide us with 
information about the chemical and physical conditions in which comets were formed. 
Importantly, following the detection of the higher than expected O$_2$ abundance in 
67P/C-G, \textit{Giotto}/Neutral Mass Spectrometer data of the coma of Oort Cloud comet 
1P/Halley have been re-analysed and an O$_2$ abundance of $3.7\pm1.7$\,\% with respect 
to H$_2$O was found \citep{RADS15},
making O$_2$ the third most abundant species in the coma.
The datasets do not compare like for like (the 67P/C-G/ROSINA observations comprise a 
time-integrated dataset, whilst the \textit{Giotto}/NMS data is a 4 hour snapsot), but
these remarkably similar O$_2$ abundances seen in comets with different dynamical histories may add 
weight to the theory that the O$_2$ is of a primordial (pre-PSN) origin and has survived 
the expected bulk thermal processing (sublimation, freezing, and re-accretion onto the comet).
Whilst the O$_2$ abundances are nearly identical, the situation for other molecular species 
is more complex; e.g.the abundances of methanol (CH$_3$OH) and carbon disulfide
(CS$_2$) are very different - both species being significantly more abundant in 1P/Halley 
\citep[see Table 3 of][]{RADS15}.
There is an interesting corollary to this; if O$_2$ is universally abundant in comets then the rate 
of its delivery to the early atmosphere of the Earth should be correlated to that of H$_2$O. The 
fact that there is little evidence of pre-biogenic O$_2$ in the atmosphere of the 
Earth could be seen as mitigating evidence against the delivery of significant H$_2$O to Earth by
comets.
% \citep{RADS15}

In addition to O$_2$, molecular nitrogen (N$_2$) was also detected (for the first time in a 
comet) in 67P/C-G (from ROSINA data) with an abundance, relative to CO, of 
(5.70$\pm$0.66)$\times 10^{-3}$. Assuming that, as in other objects with a trans-Neptunian origin, 
N$_2$ is the dominant component of the nitrogen budget this implies that N$_2$ is significantly 
depleted (by a factor of $\sim 25\times$) \citep{Rubin15}. 
As N$_2$ and O$_2$ have similar volatilities, this is also a rather surprising result.

In this paper, we address several questions:
(i) Assuming a primordial origin for the O$_2$, what does the observed abundance of O$_2$ tell
us about the physical conditions and dynamical evolution in the molecular cloud out of which 
the Solar System formed?,
(ii) What additional constraints are placed by the N$_2$ obervations in Comet 67P/C-G, and
(iii) How can these volatile ices survive the conditions in the early PSN?

In this study we revisit the possibility that the O$_2$ may indeed be primordial in nature, 
and that the O$_2$:H$_2$O ratio is effectively locked into the cometary ice through its delivery 
by the icy dust grains that agglomerate to form comets.
However, we propose that the O$_2$ is mainly formed in a transient gas-phase chemistry and then 
stored in the dust ices.
If conditions encourage the build up of ices, then they are deposited on dust grains in layers. 
If we make the gross simplification that no surface or solid-state ice-phase 
chemistry occurs then the bulk compositions of these ices represents a (time-integrated) `fossil' 
record of the evolution of the gas-phase chemistry, moderated by the varying physical conditions
(density, temperature) and freeze-out/desorption characteristics.
This means that the ice compositions are not just dependent on the instantaneous conditions in 
the molecular cloud in which they formed, they also depend on the evolutionary history of the cloud. 
In particular, the mean molecular abundances in the bulk of the ices may be very different to 
those in the gas-phase, or in the surface layers of those ices, particularly if the ices
are primarily formed during a period of transient gas-phase enhancement that is significantly 
less than the age of the cloud.

The validity of this hypothesis and the sources of uncertainty are discussed in Section~2.3 below.
However, it is a key aim of this paper to investigate what (range of) physical parameters could 
yield the observed O$_2$:H$_2$O ratio and what diagnostic power those observations may 
therefore have concerning the origins of comets. 

The nature and possible origins of the O$_2$ excess, together with our hypothesis are 
discussed in Section~\ref{sec:O2origin}. Our physical and chemical models are described 
in Section~\ref{sec:models},
and the results are presented and discussed in Section~\ref{sec:results}. In 
Section~\ref{sec:O2survival} we consider the survivability of interstellar O$_2$ ices in the PSN, 
whilst in Section~\ref{sec:N2origin} we address the observed and modelled abundances of 
N$_2$. Our key conclusions are given in Section~\ref{sec:discussion}.

\section{Possible origins of the O$_2$ excess}
\label{sec:O2origin}

\subsection{A primordial origin for the O$_2$}

If the assumption is made that the O$_2$ is primordial in nature (i.e. reflecting the 
abundances in the nascent molecular clouds from which the comets formed) then the high
O$_2$ abundances that are detected in comet 67P/C-G would seem to be at odds with both 
observations and (equilibrium) astrochemical models; which predict O$_2$ abundances 
of at least an order of magnitude lower in dark clouds than these values 
\citep[e.g.][]{WCCC16}, although there may be an obvious reason for this - as discussed above.

In a very detailed dynamical/chemical study, linked to a specific model of 
protostellar disk formation and evolution, \citet{TQ16} considered three 
possibilities:
(i) a primordial (dark cloud) origin,
(ii) formation during protostellar disk formation, and
(iii) production during luminosity outbursts in the disk.
The latter two rely on O$_2$ formation as a photolytic product of ice desorption,
followed by freeze-out in semi-porous or clathrate ices. 
The authors favour (i); the O$_2$ can be produced in primordial dark clouds and 
survive until accretion into cometary bodies, although the models require an 
exceptionally low H/O ratio ($<0.03$), a `warm' temperature of $\sim$20\,K and 
high densities ($>10^5$cm$^{-3}$) in the primordial cloud. 
Chemically, in these models, the dominant formation channels for O$_2$ are
solid-state reactions occurring in the surface layers of ices.
In the denser regions of clouds, the dominant pathways for the surface 
chemistry of O$_2$ and the formation of hydrogen oxides approximately 
simplify to:
\[ {\rm O + O/OH \to O_2} \]
\[ {\rm O_2 + H \to HO_2} \]
\[ {\rm HO_2 + H \to H_2O_2} \]
\[ {\rm H_2O_2 + H \to H_2O + OH} \]
Whilst this last reaction has a significant activation barrier in the 
gas-phase, tunnelling in (predominantly water) ices allows it to 
proceed rapidly, even at very low temperatures ($\sim$15K), via surface reactions 
\citep{Letal16}. 
Note that the first of these reactions is in competition with
\[ {\rm O + H \to OH} \]
preferentially leading to the formation of water, so that the O$:$H ratio is 
critical in defining the O$_2$ formation efficiency. In this scheme, therefore,
the relative timescales for freeze-out and free-fall have a major controlling influence.
Additionally, the warm temperatures encourage oxygen atom mobility and hence O$_2$ formation, 
whilst at the same time enhancing the sublimation of atomic hydrogen.
The O$_2$ can be further oxidised to O$_3$, which itself can then be 
hydrogenated to OH \citep[e.g.][]{ICR08}.
Thus, the surface formation of O$_2$ should be accompanied by the presence of 
HO$_2$, H$_2$O$_2$ and O$_3$.

Whilst the models do not give an obvious explanation for the low observed absence of 
N$_2$, they suggest that if the N$_2$ chemistry is somewhat slower than the 
oxygen chemistry, then it could freeze out at later times. A similar result 
could obtain if N$_2$ is slightly more volatile than CO. Both possibilities
would result in a layered ice in which the N$_2$ is not so closely associated 
with the water ice and hence is significantly more volatile, perhaps not surviving
transport into the PSN.

The models of \citet{TQ16} provide a viable mechanism that may explain the observed 
O$_2$:H$_2$O ratio as well as the good correlation between these species,
albeit for somewhat anomalous physical conditions (warmer and denser than 
is typical for a dark cloud).
This finding is only robust if the O$_2$ is formed on the primordial dust 
grains - a key result being that these ice compositions can survive into the 
forming protosolar nebula.
However, the models are somewhat less successful at explaining the low abundances of 
some other species, such as O$_3$ and HO$_2$.
The authors postulate that this could be explained by the presence of a barrier 
(of $\sim$300K) for the O+O$_2$ and H+O$_2$ reactions, effectively 
suppressing much of the O$_2$ surface chemistry. 

\subsection{Alternative origins for the O$_2$ excess}

A number of other theories have been postulated that might explain the origin of the O$_2$
(and the relative absence of O$_2$H, H$_2$O$_2$ and other species).
These are based on the idea that the O$_2$ may be formed {\em in situ.} by a variety of mechanisms,
including
% \begin{enumerate}
%\item A primordial origin in interstellar ices, driven by surface chemistry (see discussion below)
radiolysis (by high energy particles), photolysis (by solar photons), dismutation of H$_2$O$_2$, 
and high energy ion-surface reactions.
% \end{enumerate}

Radiolysis, the chemical consequence of the impact and penetration of high energy particles
into ices can result in the splitting of H$_2$O into O$_2$:
\[ {\rm 2H_2O + crp \to 2H_2 + O_2.} \]
The hydrogen atoms and molecules produced in this process would then diffuse out of the ice, so
that O$_2$ is not re-hydrogenated to H$_2$O.
However, the particles in the Solar wind have limited pentration depths and would only affect 
the outermost ice layers of the comet itself. Instead, it has been shown that the 
observed O$_2$ abundance in 67P/C-G can be produced solely via 
the radiolysis of primordial grains in the early, dense, PSN - {\em prior} to their agglomeration
into the comet \citep{Metal16b}.
Various dynamical/thermal scenarios have been considered; trapping of the O$_2$, or
sublimation, followed by re-freezing in ice `cages' or clathrates, or else as pure ices. There are a 
number of problems with this theory:
(i) the process is very slow (so the molecular cloud would have to have been very old), 
(ii) it requires all of the energy absorbed by the water ice to be used in this reaction, 
with no local heating or ionization, and 
(iii) radiolysis may produce significant amounts of H$_2$O$_2$, HO$_2$ and, most significantly, O$_3$.

A possible explanation for the relative absence of H$_2$O$_2$ is that the molecule is 
subject to a redox (dismutation/disproportionation) reaction
\[ {\rm 2H_2O_2 \to 2H_2O + O_2} \]
during the sublimation of H$_2$O ice \citep{DMB17}.
This has the advantage of by-passing the activation barrier problem 
described above, and provides a mechanism whereby O$_2$ can be produced at 
late times, when trapped H$_2$O$_2$ is desorbed.
However, the hypothesis requires efficient incorporation of H$_2$O$_2$ into 
the ice and complete conversion to O$_2$ to explain both the observed O$_2$ abundance and 
the low relative abundance of H$_2$O$_2$.

A very different suggestion \citep{YG17} is that the O$_2$ does not originate in
the cometary ices at all but, instead, is formed following the 
prompt (Eley-Rideal) reaction of energetic H$_2$O$^+$ ions with surface oxides 
(in silicates etc.). This results in the production of excited 
`oxy-water' (H$_2$O-O$^\star$) isomeric to, but the not the same as,
hydrogen peroxide. This dissociates to H$^+$ and the excited state 
HO$_2^-$ cation, which in turn dissociates to O$_2^-$. This ion is 
subject to photodetachment, yielding O$_2$.
Such a mechanism naturally explains the absence of O$_3$ and the presence of HO$_2$ whose 
abundance, although low, is still some three times larger than what is inferred for 
interstellar clouds. However, the model has yet to be analysed carefully in the context of the 
temporal correlations of the various molecular species detected by ROSINA \citep{Hass15}.

\subsection{A gas-phase primordial origin of O$_2$}

Starting from OH - either produced in gas-phase ion-neutral reactions, or via surface chemistry - 
the gas-phase chemistry of O$_2$ in dark cloud conditions is relatively simple, 
the formation being dominated by 
\[ {\rm O + OH \to O_2 + H.} \]
This reaction is in competition with the freeze-out and hydrogenation of atomic oxygen:
\[ {\rm O + grain \to ......H_2O(s.)} \] 
The main destruction channel of O$_2$ is
\[ {\rm O_2 + C \to CO + O} \]
whilst O$_2$ is also lost from the gas-phase by freeze-out into grain ice mantles:
\[ {\rm O_2 \to O_2(s.)} \] 
There are several competing reactions here and modern studies of the O$_2$-chemistry 
in interstellar cloud conditions \citep[e.g.][]{Yildiz13} have found that the depletion 
of O-atoms by freeze-out and conversion to H$_2$O and/or the destruction of O$_2$ by 
C-atoms (before CO formation is complete) yields low values of the O$_2$ abundance - 
typically an order of magnitude or more smaller than what is found in comet 67P/C-G.
However, these models also show that the O$_2$ abundance rises significantly after the 
efficient conversion of C to CO has occurred \citep{WCCC16}.
Moreover, the models also assume that the freeze-out of O-atoms and conversion to H$_2$O, 
which is fully retained in the ices is efficient. 
The gas-phase reactions are relatively fast at interstellar cloud temperatures, so 
that the abundance of O$_2$ is essentially determined by the abundances of free oxygen 
atoms (formation) and carbon atoms (destruction).
These, in turn, are controlled by the efficiency of CO formation, and also the freeze-out of the 
residual oxygen atoms (assuming that O$>$C).

There are thus three competing timescales: O to O$_2$ conversion, CO formation and O-atom 
freeze-out.
\citet{WCCC16} argue that, in dense cores where (a) CO conversion is complete and
(b) the effective rate for O-atom freeze-out may be reduced if the chemical conversion of O to OH 
and H$_2$O is inhibited due to the low abundance of atomic hydrogen; the O$_2$ abundance might be 
significantly enhanced in transient peaks; the competing timescales resulting in significant 
chemical non-linearities.
There is some evidence to suggest that large O/CO ratios may exist in 
dense, dark clouds \citep[e.g.][]{Vastel00}.

The possible existence of this transient phenomenon raises the possibility that 
O$_2$ abundances in the PSN may be significantly different to those predicted 
in models of quasi-static molecular clouds.
These differences could arise from non-linearities in the time-dependencies - perhaps enhanced 
by the dynamical history of collapsing clouds, coupled with the fact that the dust ice mantles 
have the ability to preserve a `fossil record' of gas-phase chemical evolution.
The magnitude of these deviations from equilibrium will therefore depend on a number of factors, including the
dynamics of the flow, the freeze-out and desorption characteristics of O, O$_2$ and other species,
and the physical conditions within the gas (the temperature, density, and cosmic ray ionization rate etc.). 

An additional benefit of this hypothesis is that it would help to explain the relatively
low abundances of HO$_2$, H$_2$O$_2$ and O$_3$ which are part of the same 
surface-chemistry network that produces O$_2$.

\section{The physical and chemical model}
\label{sec:models}

We have constructed a model that adopts these ideas to study how the dust ice-mantle composition 
depends on the physical and chemical evolution of the material from which comets form.
Recognising that protoplanetary and protosolar disk physical/chemical models are not at a 
sufficient maturity to identify a `standard' model of the origin and evolution of cometary bodies, we 
do not elaborate on the poorly-constrained details of this evolution, but instead
opt to implement a simple physical representation of the physical evolution, coupled with a 
more complex chemical model. This model tests the sensitivity of the chemical evolution to the dynamical 
evolution of the gas, and identifies how the abundances depends on key physical parameters, rather 
than resulting from artefacts of a detailed, specific model.

To represent the collapse and evolution of the protosolar nebula into the protosolar disk we 
therefore implement a simple model, specifically created for this study, that follows 
the chemical and physical evolution of the gas in three phases:

\begin{itemize}
\item[I.)] A static, diffuse or dark cloud phase, of duration $\sim 10^7$years, to 
simulate the chemical conditions in the interstellar cloud from which the Solar System 
formed. 
\item[II.)] A collapse phase - which we model as a free-fall collapse with a simple infall 
velocity retardation scaling factor ({\em B}, which simulates the effects of quasi-static 
contraction, magnetic braking, non-spherical effects, disk vorticity etc.).
The formulation for the free-fall contraction is as given in \citet{RHMW92}. This leads to
\item[III.)] A final, static, high density, phase which represents the conditions in the early 
Kuiper Belt/Oort Cloud.
\end{itemize}
The model follows the time-dependence of the chemistry of a single (representative) point 
as the physical conditions (density, gas and dust temperatures, extinction etc.) evolve through 
these three phases. The differential equations describing the chemistry and dynamics are 
co-integrated with the DLSODE package \citep{Hind83}.

We again emphasise that the physical properties of all three phases are very poorly constrained.
Our approach is somewhat different to the physical models adopted by other studies 
such as that of \citet{TQ16}. Although those models include considerable physical complexity, the initial conditions (prior to dynamical collapse and disk evolution) are modelled by steady-state 
evolution of a cloud for a time equivalent to the free-fall timescale. 
This will have considerable bearing on the assumed composition of the ices and,
unsurprisingly, that approach yields very low (quasi-equilibrium) 
values for the O$_2$ abundances in both the gas-phase and in the ices.

However, as explained above, the ices effectively contain a fossil record of the dynamical and 
chemical evolution and the relative contributions of the chemistry in each of the three dynamical 
phases will depend on the competitions between relevant timescales, such as those for gas-phase 
chemistry, freeze-out/desorption and dynamical evolution. 

In those situations where the conditions in Phase I are equivalent to those of a diffuse cloud, 
freeze-out of icy mantles will not occur until that time in Phase II when the density has risen 
(and the dust temperature fallen) such that thermal desorption is quenched.   
In Phase II we simply assume homologous collapse, in which case the extinction (A$_v$) scales as 
$n^{2/3}$, yielding very large extinctions by the beginning of phase III. The dependency of 
$T_{gas}$ and $T_{dust}$ on A$_v$ is taken from semi-empirical modelling studies 
\citep[e.g.][]{KC10}, constrained by the boundary conditions imposed by the specified 
temperatures in Phases I and III. 
The mean dust grain radius and surface area per hydrogen nucleon (given in Table 1) imply a normal interstellar dust to gas ratio.

In Figure~1 we show the evolution of the density, extinction and 
the temperatures of the gas and dust components in the collapse phase (Phase II).
\begin{figure*}
\label{fig:PhaseII}
\begin{center}
\includegraphics[width=10cm]{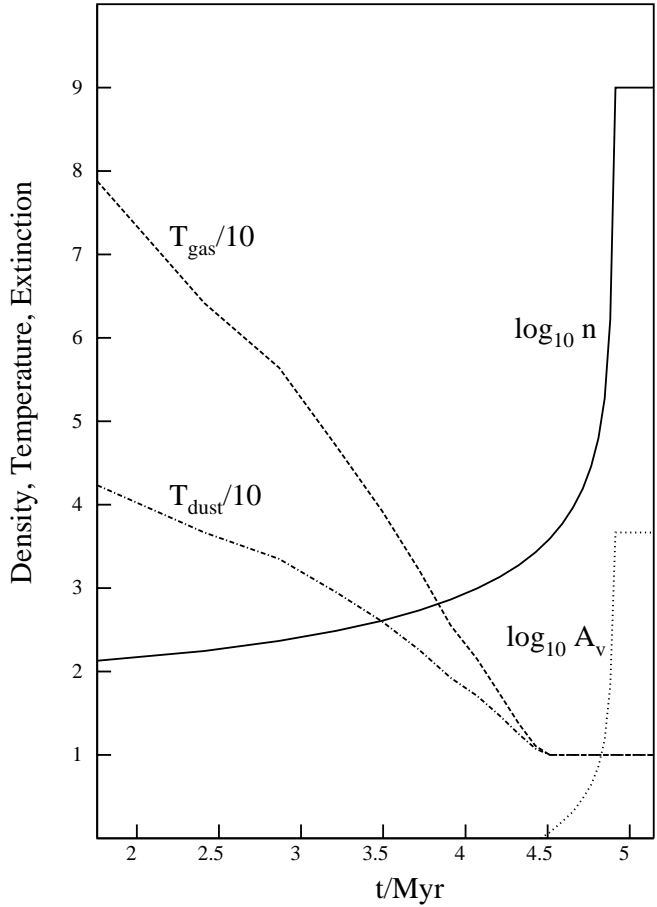}
\caption{The time-evolution of the density, extinction, and the gas and dust 
temperatures in the collapse phase (Phase II.)}
\end{center}
\end{figure*}

It seems likely that the cosmic ray flux will be modulated (reduced) due to exclusion by young stellar 
winds \citep{CLA14} and/or magnetic field channelling \citep{FA14} in the early PSN, and the 
assumption that typical (interstellar) cosmic ray ionization fluxes apply throughout the 
protostellar/protoplanetary evolution is almost certainly inaccurate.
We have therefore considered a range of cosmic ray ionization rates in our models.
However, we found that the results are insensitive to the time-dependence of the cosmic ray ionization rate during the collapse pahse so, for the sake of simplicity and lacking any empirical
constraints, we make the simpliying assumption that it is constant throughout all three phases.

Our chemistry includes a comprehensive network of 1296 neutral-neutral and ion-neutral gas-phase 
reactions and 134 gas-grain (freeze-out and desorption) and surface reactions, between 96 gas-phase 
chemical species  and 22 solid-state species. This gives good coverage of the formation and destruction channels for O$_2$. The relevant oxygen-bearing species in the solid state include: O, O$_2$, OH, H$_2$O, CO, and CO$_2$.

We use a highly simplified representation of the surface chemistry, so that we can differentiate 
between the gas-phase processes that we are studying, and the surface reactions leading to O$_2$ 
formation and destruction, that have been reported elsewhere.

Thus, whilst we include the surface oxidations and hydrogenation of O, O$^+$, OH, OH$^+$, 
O$_2$ to oxygen hydrides, O$_3$, and (in reactions with CO) to CO$_2$, the rate coefficients are set to zero for many of these reactions in our standard model.
We do, however, assume efficient hydrogenation of O and OH, so that all O and OH 
species that impact and stick to grains are fully hydrogenated to H$_2$O ice. 
We further assume that all H$_2$O so-formed remains in the ice.
We do not allow for any conversion of surface O or OH to O$_2$ (possibly underestimating the 
O$_2$:H$_2$O ratio), but nor do we allow 
for the hydrogenation or oxidation of surface O$_2$ to products such as O$_3$, 
and leading to H$_2$O formation. Obviously, these processes would lead to lower O$_2$:H$_2$O ratios.
However, we also note that there is a notable deficiency of H$_2$O$_2$ in the coma of comet 67P/C-G
so that it would seem that the sublimation conversion of O$_2$ to H$_2$O via H$_2$O$_2$ \citep{DMB17}
may not be efficient in these conditions. We have therefore chosen not to include this channel - 
but recognise that its omission may result in our overestimating the O$_2$:H$_2$O ratio. 

We also allow for surface reactions of CO with O and OH to form CO$_2$ with 10\% efficiency, 
yielding the empirical CO:CO$_2$ ratio of $\sim 1$ \citep[e.g.][]{Chiar11}.
As dictated by laboratory experiments and the observed abundances of complex organic molecules 
in astrophysical sources, surface CO is assumed to convert to CH$_3$OH and H$_2$CO with an efficiency of 
$\sim$ 10\% \citep{WK02}.    

We have included a full range of thermal and non-thermal desorption mechanisms, including; photodesorption, cosmic-ray induced photodesorption, cosmic ray heating 
(hot-spot and whole grain) and formation-enthalpy driven desorption. 
The desorption that is induced by cosmic ray heating results from the bulk heating of
the ice mantles, or even the whole grains. However, photodesorption (due to to both the primary
interstellar radiation field and that induced by cosmic rays) only affects the surface layers
of the ices.
To ensure that these processes are correctly modelled, we determine the chemical composition of 
the ices on a layer-by-layer basis, and consider the desorption efficiencies for each layer.
We adopt commonly used values for the the yields for the various desorption processes 
\citep[e.g.][]{HKBM09} although, 
in our standard model, we assume that the yield for H$_2$-formation-induced desorption is zero, 
in line with results from models of starless cores (Rawlings, Keto \& Caselli 2019).

As the nature of the cometary ices is not necessarily similar to that which exists in
interstellar dark clouds and the roles of photolysis/radiolysis are not clear we consider 
that the inclusion of more elaborate surface chemistries (e.g. to form complex organic 
molecules) is not justified in this study.

The standard physical parameters, together with key chemical parameters (abundances and binding energies) that we use in this model are summarised in Table~1.
The chemical initial conditions in the diffuse gas (Phase I) are taken to be atomic 
+ H$_2$. The model results are not sensitive to variations in this assumption.

\begin{table}
\centering
\caption{Physical and chemical parameters for the standard model.}
\label{tab:param}
\begin{tabular}{|l|l|l|}
\hline
 Parameter & Symbol & Value \\
\hline
 Phase I density & $n_I$ & 100 cm$^{-3}$ \\
 Phase I gas temperature & $T_{g,I}$ & 100 K \\
 Phase I dust temperature & $T_{d,I}$ & 50 K \\
 Phase I extinction & $A_{v,I}$ & 0.1 magnitudes \\
\hline
 Phase II free-fall retardation factor & B & 1.0 \\
\hline
 Phase III density & $n_{III}$ & 10$^{9}$ cm$^{-3}$ \\
 Phase III gas temperature & $T_{g,III}$ & 10 K \\
 Phase III dust temperature & $T_{d,III}$ & 10 K \\
\hline
 Cosmic ray ionization rate & $\zeta_0$ & $1.3\times 10^{-17}$ s$^{-1}$ \\
 Mean grain radius & $<a>$ & 83\AA \\
 Mean grain surface area/H-nucleon & $<\sigma>$ & $8\times 10^{-21}$ cm$^2$ \\ 
\hline
 Helium abundance & He/H & 0.1 \\
 Carbon abundance & C/H & 2.55$\times 10^{-4}$ \\
 Nitrogen abundance & N/H & 6.1$\times 10^{-5}$ \\
 Oxygen abundance & O/H & 4.58$\times 10^{-4}$ \\
 Abundance of S, Si and Na & S,Si,Na/H & $1.0\times 10^{-7}$ \\
\hline 
 CO binding energy & E$_{CO}$/k & 960 K \\
 H$_2$O binding energy & E$_{H_2O}$/k & 5770 K \\
 O$_2$ binding energy & E$_{O_2}$/k & 1210 K \\
 N$_2$ binding energy & E$_{N_2}$/k & 710 K \\
\hline
\end{tabular}
\end{table}
\null

\subsection{Observational constraints}

The rather simplistic assumption that measurements made with ROSINA, and other 
instruments on the Rosetta mission, represent the mean abundance ratios 
throughout the coma has subsequently given way to the realisation that there 
are significant diurnal variations (due to the rotation of the comet) - 
essentially deriving from variations in illumination conditions \citep[e.g.][]{Retal17}, seasonal
variations (i.e. resulting from the tilt in the comet's spin axis), and 
variations due to morphological and compositional variations over the comet's 
surface (e.g. differences exist between the composition and outflow properties 
of the gas outgassing from the neck and the poles of the two lobes; \citealp{Fou16}). 
Morever, there are probably significant variations in the ice compositions as a 
function of depth \citep{Betal16}. However, interestingly, observations with {\em VIRTIS}
show that the refractory compositions of both lobes are similar, suggesting that the
cometesimals formed in the same region \citep{Capetal15}.

In this section we briefly summarise the observational data for comet 67P/C-G and identify the
constraints that we use to test the success of our model:

The O$_2$:H$_2$O ratio was measured by ROSINA to be 3.80$\pm$0.85\% (and a similar value was 
obtained for comet 1P/Halley: O$_2$/H$_2$O=3.7$\pm$1.7\%, \citealp{RADS15}). The 
variations in the O$_2$ and H$_2$O are fairly well-correlated and, in addition, the O$_2$ distribution
is approximately isotropic and does not vary significantly with time or with heliocentric 
distance \citep{Bieler15}.
The same study also found low values for the abundances of other oxygen hydrides;
HO$_2$:O$_2$ = 0.19\%, H$_2$O$_2$:O$_2$ = 0.06\%, and also O$_3$:O$_2 <$0.0025\%.

ROSINA observations yielded an N$_2$:CO ratio of $\sim$0.17-1.6\%, with a mean of 0.57\%. The two
species, of similar volatility, were found to be only moderately correlated, and vary with position 
above the surface of the comet. The N$_2$ and CO are poorly correlated with H$_2$O \citep{Bieler15}.
Because of this, and the fact that the H$_2$O production rate may change with heliocentric distance,
\citet{Rubin15} consider the N$_2$:CO ratio to be a better metric of the N$_2$ abundance than the 
N$_2$:H$_2$O ratio.
Making the reasonable assumption that N$_2$ and CO are the dominant reservoirs of nitrogen and 
carbon in the PSN and that the conversion of N to N$_2$ is essentially complete \citep{Rubin15}, 
the N$_2$:CO ratio should reflect the elemental abundances of N and C in the PSN, yielding an 
expected (undepleted) ratio of 14.5\%. 
The observed value therefore suggests that N$_2$ is depleted by a factor of $\sim 25\times$ \citep{Rubin15}. In addition
\citet{Bal15} found that $^{36}$Ar is well-correlated to N$_2$ (which has a similar volatility) 
so that $^{36}$Ar:N$_2$ = 0.9$\pm$0.03\%. The poor correlation of both species with H$_2$O means 
that the inferred N$_2$:H$_2$O ratio is somewhat less well-constrained at 0.011-0.26\%.

The relationship between CO/CO$_2$ and H$_2$O is very much more complex and ill-defined. Whilst 
CO:H$_2$O = 10-30\% over the sunlit hemisphere \citep{Rubin15}, the CO:H$_2$O and CO$_2$:H$_2$O 
ratios are highly variable; qualitatively they are reasonably well-correlated, but there are 
huge variations in the 
abundance ratios, both diurnal and seasonal/morphological, so that the ratios can exceed unity.
Thus CO/H$_2$O ranges from 0.13($\pm$0.07) to 4($\pm$1), and CO$_2$/H$_2$O ranges from 0.08($\pm$0.05) to 8($\pm$2).
There is less variation in the CO and CO$_2$ abundances than in that of H$_2$O, and the 
H$_2$O peaks when oberving the neck, whilst the CO$_2$ peaks when observing the underside of the body
of the larger of the two lobes \citep{Hass15}. The surface CO:CO$_2$ ratio was also measured to be 
$\sim$ 0.07 with Ptolemy (the mass spectrometer aboard the Philae lander), significantly smaller 
than the values inferred in the coma from ROSINA measurements ($\sim$ 0.50-1.62) \citep{Betal16}.
Taken together, these facts may indicate that there are possible compositional inhomogeneities and 
that the CO and CO$_2$ sublimate from greater depths in the ice than the H$_2$O, although the 
possibility remains that these could simply be the result of variations in illumination. In their 
anaylsis, \citet{Betal16} found that there was no clear distinction as to whether the H$_2$O ice is primarily 
crystalline, amorphous, or composed of clathrates; to which we return in section~\ref{sec:O2survival}
below.

Finally, we note that other more complex organic molecules (some of which are nitrogen-bearing) 
have also been detected.
Many sulfur-bearing molecules were also detected by ROSINA \citep{Caletal16},
although they are notably absent in COSAC measurements \citep{Getal15}.
As our understanding of sulfur chemistry (and particularly surface reactions) is very
poor, we have not attempted to model the sulfur chemistry.
To summarize the data that are relevant to this study, in Table~2 we specify 
the observational constraints that we use to test our model.
Note that we do not include an accurate representation of the surface chemistry of
HO$_2$, H$_2$O$_2$ and O$_3$ and so the observed abundances of these species are not used to
constrain our model.

\begin{table}
\centering
\caption{Observational constraints on abundance ratios in Comet 67P/C-G.}
\label{tab:constraints}
\begin{tabular}{|l|l|l|}
\hline
 Parameter & Nominal value & Notes \\
\hline
 O$_2$:H$_2$O & 3.8\% & Well-correlated (1-10\%) \\
 HO$_2$:O$_2$ & 0.19\% & $\sim 3\times$ interstellar value \\
 H$_2$O$_2$:O$_2$ & 0.06\% & $\sim$ interstellar value \\
 O$_3$:O$_2$  & $<$0.0025\% & Non-detection (upper limit) \\
\hline
 N$_2$:CO & 0.57\% & Moderate correlation (0.17-1.6\%) \\
 N$_2$:H$_2$O & 0.011-0.26\% & - \\
\hline
 CO:H$_2$O & 13-400\% & Coma, general, highly variable \\
           & 10-30\% & Sunlit hemisphere \\
 CO$_2$:H$_2$O & 8-800\% & Highly variable \\
 CO:CO$_2$ & 50-162\% & Coma \\
           & 7\% & Surface \\
\hline
\end{tabular}
\end{table}
\null

\section{Results}
\label{sec:results}

We initially applied a simple (static) model to perform an analysis that allowed 
us to determine which physical parameters are the most significant in determining 
the O$_2$:H$_2$O ratio. 
We found that enhanced O$_2$:H$_2$O ratios (of the order of 1\%) are obtained if 
the number density lies in the range $10^4-10^5$cm$^{-3}$. The abundance of O$_2$ 
is also enhanced if we make the assumption that a fraction ($\sim$10\%) of the 
H$_2$O that is formed by surface reactions on dust grains is desorbed via the 
enthalpy of formation of the molecule.

% (in models assuming no O, OH or CO is converted to CO$_2$, nor any enthalpy-driven desorption).... 

In the density range; $n\sim 10^4-10^5$cm$^{-3}$, the O$_2$:H$_2$O ratio is found 
to be particularly sensitive to the cosmic ray ionization rate, primarily due to 
the relatively high efficiency of O$_2$ desorption by cosmic-ray heating of grain mantles. 
If $\zeta$ is as low as $10^{-18}$ s$^{-1}$, then
O$_2$:H$_2$O may be as high as $\sim 10$\% in equilibrium. 

From this initial analysis, we deduced that the key free parameters are; the density 
and temperature in phase III, the free-fall collapse retardation factor ($B$) and 
the cosmic-ray ionization rate ($\zeta$). 
The range of parameters that we have investigated in the full model are: 
$B=0.1-1$, $n_{\rm III}$=10$^7$-10$^{11}$cm$^{-3}$, $T_{\rm III}$=10-20\,K, 
$\zeta$/$\zeta_0$=0.1-10.0 ($\zeta_0=1.3\times 10^{17}$s$^{-1}$), and 
$G/G_0$ = 0.5-10.0 (where G$_0$ is the standard interstellar radiation field).

We have considered all possible parameter combinations and present a selection of 
representative results in Figure~2 and Table~3.

In Figure~2 we show the time-evolution of the abundances of relevant gas-phase and 
ice-phase species in the 
collapse phase (Phase II) and the final, steady state phase (Phase III).
The figure shows many of the key features of the chemical behaviours. Thus, there is a clearly visible 
threshold density for the onset of ice formation, above which the extinction is sufficiently high and the dust
temperature drops sufficiently low that the rates for accretion exceed those for desorption. 
The effects of the non-linearities in the chemistry are also visible in the transient behaviours of species such as O$_2$
and N$_2$; the abundances of which show strong peaks in the collapse phase, but then decline at later times.
It is worth noting (see section~\ref{sec:N2origin} below) that there is very little 
qualitative difference between the evolution of the O$_2$ and N$_2$ abundances.
It is also evident that the non-linear behaviour is limited to the collapse phase (Phase II); 
there is very little chemical evolution in the PSN phase (Phase III), where the main chemical processes are surface reactions (e.g. the gradual conversion of CO to CO$_2$) and a slow response to the changing balance between the freeze-out and desorption processes.   
Comparing the evolution of the abundances in Figure 2 to the dynamical evolution in Figure 1
it can be seen how the period of O$_2$ enhancement is relatively short ($<0.5$Myr) compared to 
the dynamical age of the cloud but, crucially, occurs at the time of bulk ice mantle accretion.
The ices therefore contain abundances of O$_2$ that are significantly larger than stochastically
observed gas-phase abundances in dark clouds.

The essential numerical output from the model consists of the abundance ratios of key species at 
snapshots in time.
In Table~3 we give the values of the (solid state) abundance ratios of
O$_2$:H$_2$O, N$_2$:H$_2$O, N$_2$:CO, and CO:H$_2$O 
(a) at the end of the collapse phase (Phase II), and
(b) after 100\,Myrs evolution in the final, dense phase (Phase III). 
From this table we can see some of the sensitivities of these ratios to the various free parameters but,
overall, we can see that the modelled abundance ratios are fairly close to the observationally inferred
values in most cases and are quite robust to variations in the values of the free parameters.
Specifically, the O$_2$:H$_2$O, N$_2$:H$_2$O, N$_2$:CO and CO:H$_2$O ratios typically lie in the ranges of
1-3.7\%, 5-13\%, 5-17\%, and 100-142\% respectively. In a few cases (e.g. for $n_{\rm III}$=10$^7$cm$^{-3}$
or $\zeta/\zeta_0=10$) there is a strong late-time evolution of the N$_2$:CO and CO:H$_2$O ratios, which
is primarily driven by long-term desorption effects.

% Also relevant are: 
% (a) the proportion of the ice that is composed of H$_2$O, 
% (b) the proportion of molecular material that exists in the solid-sate, and 
% (c) the gas-to-ice ratio of H$_2$O.

The important quantitative results that we have obtained from the three-phase model are:
\begin{itemize}
\item A good fit to the O$_2$:H$_2$O ratio (2-5\%) is obtained whenever $B=0.1$ and 
$\zeta/\zeta_0<10$, regardless of the values of $n_{\rm III}$ and $T_{\rm III}$. A lower, but 
reasonable ratio $\sim$ 1-2\% is obtained for $B=1.0$.
\item The modelled N$_2$:CO abundance ratio is closest to the observed 
values (0.17-1.6\%) when the collapse is slow ($B=0.1$). 
Values of 1.5-2\% can be obtained in somewhat extreme conditions
($n_{\rm III}=10^7$cm$^{-3}$, $\zeta/\zeta_0 = 10$ and $T_{\rm g,III}=20$K) but these also 
result in low O$_2$:H$_2$O ratios of $\sim 0.1-0.2$\%. 
At higher densities ($n_{\rm III}\geq 10^8$cm$^{-3}$) and when $\zeta/\zeta_0 = 0.1$ 
and $T_{\rm g,III}=10$K, slightly larger values ($\sim 2.7-3.1$\%) are obtained  
but these models also yield O$_2$:H$_2$O $\sim 3.1-3.3$\%.
In general, the ratio is found to be very sensitive to the cosmic ray ionization rate, 
so that for $\zeta/\zeta_0 = 10$, desorption leads to CO-poor ices and very large 
values result. 
% The best fits (N$_2$:CO=1-3\%) are obtained for $B=0.1$, $\zeta/\zeta_0 = 0.1$ and $T_{\rm III}$=10K.
\item N$_2$:H$_2$O ratios of $\sim$4-15\% are produced in most of the models, considerably larger than 
the observed values. A good fit can be obtained for low values of n$_{\rm III}$, coupled with high values
of $\zeta$; especially at late times - but these largely result from the effect of 
efficient desorption, which also leads to very low O$_2$ abundances. However, an 
O$_2$:H$_2$O ratio of $\sim$1\% can be obtained if G/G$_0$=1 and B=1.0.
\item The modelled CO:H$_2$O abundance ratio is typically very large (of order unity) which is 
compatible with observations. Values of $\sim$10-40\% (as measured for the sunlit hemisphere) 
can be reproduced for low values of n$_{\rm III}$ (10$^7$-10$^{8}$cm$^{-3}$), and $\zeta/\zeta_0 = 1$. 
This ratio is also very sensitive to $\zeta/\zeta_0$, being strongly suppressed for high cosmic 
ray ionization rates.
\end{itemize}

Taken together, the best fits for the O$_2$:H$_2$O and N$_2$:CO ratios are obtained for
$B=0.1$, $\zeta/\zeta_0 = 0.1$, $T_{\rm III}$=10K and any value of $n_{\rm III}$. For these parameters 
CO/H$_2$O$\sim$1.

We therefore find that, for a reasonably large volume of the parameter space that we have 
investigated, we can obtain a good match to the observed values of the O$_2$:H$_2$O ratio and a resonable match to the observed N$_2$:CO ratio.
A good match to the N$_2$:H$_2$O ratio can only be obtained with a more contrived parameter combination,
and most of our models yield CO:H$_2$O abundance ratios of order unity, comparable to the 
obserationally inferred values.

% Other runs/parameter variations we should consider/comment on:
% \begin{itemize}
% \item Start from dark cloud conditions ($n_I$=10$^4$cm$^{-3}$, $T_{g,I}$=T$_{d,I}$=10K, A$_v$=?)
% \item (Results are insensitive to the threshold temperature for gas-grain 
% interactions (27K, 49.9K tried) - but vary relative binding energy of N$_2$ ?)
% \item Try T$_{gas,III}$=T$_{dust,III}$=30K (may need to change ATOL/RTOL)
% \item Try G$_0>10$
% \item Higher values of $\sigma_H$ (give larger values of all ratios)
% \end{itemize}

\begin{figure*}
\label{fig:res}
\begin{center}
\includegraphics[width=12cm]{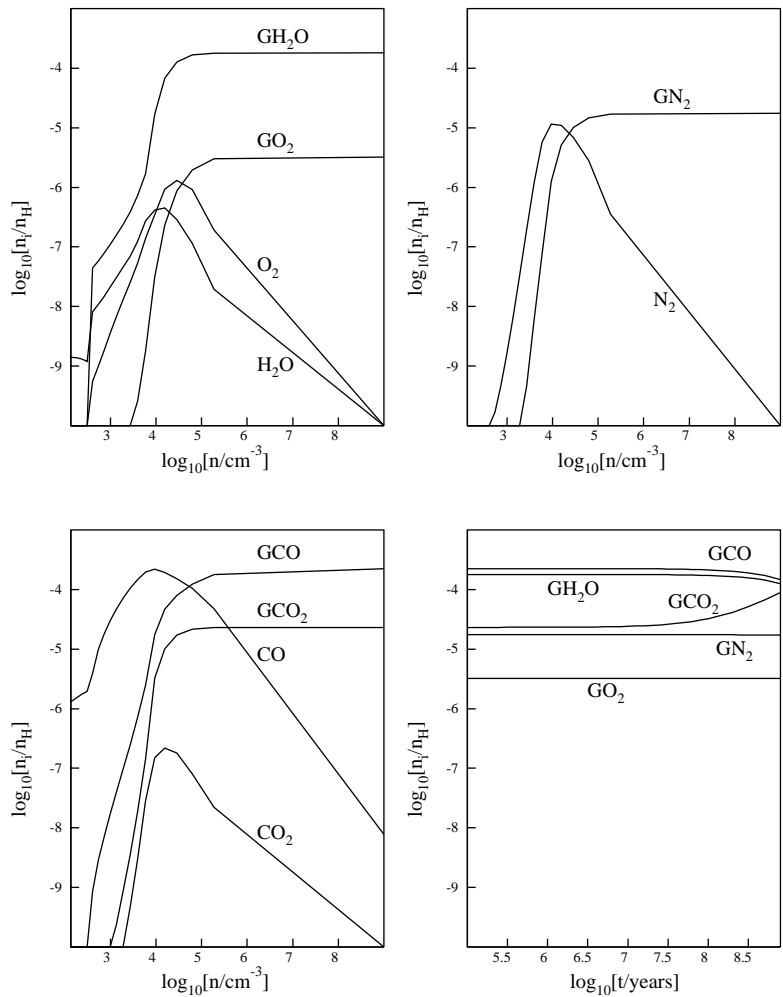}
\caption{The time-evolution of the fractional abundances (relative to hydrogen nucleons) 
of key molecular species. Solid-state species are 
prefixed by a `G' to indicate the ice abundances on `grains'. The dependences on density 
(in the collapse phase; Phase II) are shown in the top left, top right and bottom left frames.
The bottom right frame shows the time-dependences of the ice species in the final, static phase (Phase III.)}
\end{center}
\end{figure*}

\begin{table*}
\label{tab:res}
\centering
\caption{Results from the model. The cosmic ray ionization rate is given relative to the 
standard value of $\zeta_0=1.3\times 10^{-17}$ s$^{-1}$. In all of these models:
$n_{\rm I}$=100cm$^{-3}$, $T_{\rm I,g}$=100K and $A_{\rm v,I}$=0.1.
Columns 6-13 give the values of the abundance ratios (a) at the end of the collapse phase (Phase II),
and (b) at 10$^8$ years into the final steady-state phase (Phase III).}
\label{tab:results}
\begin{tabular}{|l|l|l|l|l|l|l|l|l|l|l|l|l|}
\hline
$B$ & $n_{\rm III}$ & $T_{\rm III}$ & $\zeta/\zeta_0$ & $G_0$ & 
\multicolumn{2}{l}{O$_2$:H$_2$O} & \multicolumn{2}{l}{N$_2$:H$_2$O} & \multicolumn{2}{l}{N$_2$:CO} & \multicolumn{2}{l}{CO:H$_2$O} \\
 &  &  &  &  & a & b & a & b & a & b & a & b \\
\hline
1.0 & 10$^9$ & 10 & 1.0 & 1.0 & 1.78\% & 1.86\% & 9.70\% & 10.10\% & 7.77\% & 8.15\% & 124\% & 123\% \\ 
0.1 & 10$^9$ & 10 & 1.0 & 1.0 & 2.94\% & 3.07\% & 10.66\% & 11.12\% & 10.08\% & 10.67\% & 105\% & 104\% \\
\hline
1.0 & 10$^7$ & 10 & 1.0 & 1.0 & 1.78\% & 1.17\% & 9.68\% & 5.52\% & 7.80\% & 17.23\% & 124\% & 32\% \\ 
1.0 & 10$^8$ & 10 & 1.0 & 1.0 & 1.78\% & 1.72\% & 9.70\% & 9.20\% & 7.77\% & 8.73\% & 124\% & 105\% \\ 
1.0 & 10$^{10}$ & 10 & 1.0 & 1.0 & 1.78\% & 1.88\% & 9.70\% & 10.20\% & 7.76\% & 8.10\% & 125\% & 126\% \\ 
1.0 & 10$^{11}$ & 10 & 1.0 & 1.0 & 1.78\% & 1.88\% & 9.70\% & 10.21\% & 7.76\% & 8.09\% & 125\% & 126\% \\ 
\hline
1.0 & 10$^9$ & 20 & 1.0 & 1.0 & 1.67\% & 1.75\% & 8.81\% & 9.20\% & 7.16\% & 7.50\% & 123\% & 123\% \\
\hline
1.0 & 10$^9$ & 10 & 0.1 & 1.0 & 0.80\% & 0.81\% & 7.77\% & 7.81\% & 6.19\% & 6.22\% & 125\% & 125\% \\ 
1.0 & 10$^9$ & 10 & 10.0 & 1.0 & 1.89\% & 2.18\% & 13.41\% & 12.98\% & 13.42\% & 139\% & 100\% & 9.34\% \\
\hline
1.0 & 10$^9$ & 10 & 1.0 & 0.5 & 1.90\% & 1.98\% & 8.89\% & 9.26\% & 7.08\% & 7.43\% & 126\% & 125\% \\ 
1.0 & 10$^9$ & 10 & 1.0 & 10.0 & 1.46\% & 1.53\% & 12.19\% & 12.70\% & 9.99\% & 10.51\% & 122\% & 121\% \\ 
\hline
0.1 & 10$^9$ & 10 & 0.1 & 1.0 & 3.11\% & 3.13\% & 4.31\% & 4.34\% & 3.08\% & 3.10\% & 139\% & 140\% \\ 
0.1 & 10$^9$ & 20 & 0.1 & 1.0 & 3.72\% & 3.74\% & 6.49\% & 6.52\% & 4.58\% & 4.59\% & 142\% & 142\% \\ 
0.1 & 10$^9$ & 10 & 0.1 & 10.0 & 3.09\% & 3.10\% & 6.88\% & 6.92\% & 4.92\% & 4.94\% & 140\% & 140\% \\ 
0.1 & 10$^{11}$ & 10 & 0.1 & 0.5 & 3.14\% & 3.15\% & 3.76\% & 3.78\% & 2.69\% & 2.70\% & 140\% & 140\% \\ 
0.1 & 10$^7$ & 20 & 10.0 & 0.5 & 0.31\% & 0.10\% & 3.97\% & 0.00\% & $>$20\% & 1.70\% & 0.15\% & 0.03\% \\ 
\hline
\end{tabular}
\end{table*}

\section{The survival of O$_2$ in the PSN}
\label{sec:O2survival}

Having determined that O$_2$ can be formed in the gas-phase and retained in ice mantles, we 
now address the issue of whether or not this can be incorporated, without substantial modification
due to thermal processing, into the young PSN. 
There is some debate as to the efficiency of the delivery of primordial 
volatiles from molecular clouds into the protosolar nebula, although models show that the 
abundance of H$_2$O and the HDO:H$_2$O abundance ratio are preserved,
indicating that interstellar water ice is efficiently delivered to 
protoplanetary disks \citep{Furuya17}.

To answer this question we need to know both how volatile species are trapped in the more
abundant water ices and how they respond to thermal processing.
The phases and morphology of water ice are extremely complex but, for the purpose of our discussion, we 
can identify three ways in which (mixed-in) volatiles can be retained in water ices:
(i) in crystalline water ice, which can only include very small fractions of volatiles, such as
O$_2$,
(ii) in amorphous water ice, which can hold signficant amounts of volatiles, but which is 
very susceptible to thermal processing, and 
(iii) in closed cages of water molecules, or `clathrates'.

Previous studies of the chemistry of (primordial) cometary ices have tended to ignore the 
processing of the pre-cometary material by the Sun's protostellar precursor, whose luminosity 
at the top of the Hayashi track was very much greater than that of the present-day Sun.

In this context, \citet{Metal16a} identify two possible reservoirs for ices and trapped O$_2$:
(i) in the 5-30au range; where interstellar ices have been sublimated and re-frozen
(as the PSN cooled) into crystalline and clathrate ices, and 
(ii) in the outer regions ($>$30au); in which interstellar ices are pristine, amorphous and 
semi-porous. Unfortunately the boundary between the two is poorly determined so that there is 
some ambiguity as to which source dominates in Jupiter-family comets, such as 67P/C-G, 
which originated in the trans-Neptunian Kuiper belt. 
\citet{Metal16a} favour (i) as the best way of explaining the N$_2$ and Ar abundances in 67P/C-G; with 
clathrate formation on grains at $\sim$44-50K in the cooling PSN; the comet forming from
the agglomeration of these grains, with any untrapped volatiles frozen out as pure
crystalline ices.

The efficiency with which volatile species, such as O$_2$, can be retained in 
ices and their susceptibility to desorption is thus obviously more than a 
simple dependence on the binding energies.
In general, it depends very strongly on the degree of mixing with the water ice 
and the solid-state phase of the ice itself as indicated above.
Experimentally, studies have shown that, as icy grains are gradually heated, species 
such as O$_2$ and N$_2$ can desorb in four distinct bands;
(i) from the (multi-layered) pure species (at $T\sim 20-30$K),
(ii) from a monolayer on the water ice (at $T\sim 40-50$K),
(iii) resulting from the amorphous to crystalline phase transition in the water
ice, that releases volatiles trapped in closed pore clathrates and is 
exothermic (at $T\sim 140$K), and 
(iv) co-desorption of the the remaining, trapped, volatiles with the sublimation
of the water ice itself (at $T\sim 160$K) \citep{Cetal04,Vetal04}. 
This desorption band structure is a consequence of a rather simplified unmixed/`onion-skin' 
representation of the layering of the ices with the amorphous water ice being 
`doped' by thermal diffusion from surrounding apolar ice layers.
Such a stratified structure is probably not representative of cometary ice grains, 
and also does not consider the possibility of volatile species formation as a 
result of internal solid-state reactions. 
In these discussions we must also be very careful to remember that the comet is 
best represented as an aggregate of icy grains and not to conceptualize it as a `giant grain' 
with distinct ice layers.

However, the key findings of these desorption experiments have been 
confirmed by more recent studies \citep[e.g.][]{Cetal15}. Specifically; 
(i) O$_2$ and N$_2$ have very similar desorption characteristics and will compete for
available binding sites in the ice, and (ii) water ice has the ability to both 
trap the volatile species in closed-pore clathrates (formed in the phase change 
from a more porous amorphous state, that occurs at $T\sim 30-88$K) and semi-porous ice, 
and this hugely modifies the binding property of the volatiles themselves.
It is therefore evident that the desorption properties are critically dependent on 
the way in which the volatiles are deposited and mixed with the water ice.
For co-deposited ices, only the crystallisation and co-desorption bands are relevant, 
with the crystallisation desorption accounting for up to 67\% of the total volatile 
desorption.
If the bulk ice is formed from the accumulation of very small interstellar grains, 
then it may well have a porous structure, so that most of the O$_2$ could be trapped 
in clathrates, to be released in the amorphous to crystalline phase change.

This phase change is very rapid at high temperatures ($\geq$ 130K), 
but even at temperatures as low as 90K, the transition only takes $\sim$10$^3$years,
with $T\leq$80K for long-term stability. Such conditions are unlikely in the early PSN. 
As well as stellar radiation, the phase change can also be effected by micrometeoritic
impact heating and internal radiogenic heating etc. Here it is worth noting that studies 
have shown that both crystalline and amorphous water ice are present in 
trans-Neptunian objects \citep{Tetal16}.
 
Indeed, it has been speculated that exothermic transition of amorphous to 
crystalline ice may be a significant energy source that drives outgassing from
comet 67P/C-G \citep{Agar17}.

% ...DO WE WANT TO SAY ANYTHING ELSE HERE...?

% In the simple 'onion skin' model, in which polar and apolar ices are clearly 
% separated, the fraction of the O$_2$ and N$_2$ that is released in each of the 
% four bands is $\sim$35\%, 45.5\%, 13\%, and 6.5\% (from Viti et al., 2004)

\section{N$_2$ in Comet 67P/C-G}
\label{sec:N2origin}

The apparent absence of N$_2$ in 67P/C-G is puzzling, and is not well-reproduced by 
our models.
N$_2$ has similar desorption characteristics to CO; the desorption energies of
both species are highly dependent on the degree of interaction with water
ice, but - irrespective of environment - the ratio of the desorption energies
for N$_2$ and CO is $\sim$0.9, implying that the desorption temperatures are 
within a few Kelvin of each other \citep{Fetal16}.

As with the chemistry of O$_2$, N$_2$ formation competes with the hydrogenation of
nitrogen to NH$_3$ - both in the gas-phase and on the surface of grains. The chemistry of 
N$_2$ is somewhat different and less well understood than that of O$_2$. In our models we 
find that, typically, the N$_2$:NH$_3$ ratios are of order unity, although there are 
significant margins of error in these values.

Another possibility is that the strong dependence of the binding energy on the presence 
of water implies that the desorption properties of N$_2$ and CO would be very different if, 
for example, the N$_2$ predominantly resides in a less water-rich environment than the CO;
in which case the N$_2$ would effectively be much more volatile.
However, our modelling shows little evidence for significant differences in the 
time-dependencies of the chemistry of O$_2$ and N$_2$ which would be necessary to explain 
this layering.

Alternatively, laboratory experiments show that N$_2$ may be inefficiently trapped in 
water ices as compared to CO, to a degree that could explain the observed N$_2$/CO ratio. 
Thus \citet{Rubin15} speculate that N$_2$ is inefficiently trapped above $\sim$24K, and
it may also be inefficiently trapped in clathrates.

However, the observed strong correlation between Argon and N$_2$ is consistent with both 
species being included in ices at very low temperatures \citep{Bal15}.
A possibility is that the N$_2$ could have been trapped more efficiently at lower temperatures 
($\sim 20$K) but then subject to efficient thermal sublimation in the high luminosity 
protostellar phase. 

Other mechanisms that may result in suppressed N$_2$ abundances include
radiogenic heating due to the decay of Aluminium and Iron isotopes, which could result 
in the desorption of volatiles, such as N$_2$, Ar and CO, during the accretion phase. 
Indeed, the available energy may be sufficient to result in the conversion of
simple nitrogen-bearing species, such as N$_2$, HCN, and NH$_3$, into complex 
organics and possibly even amino acids \citep{Metal17}.

Finally, we should note that the nitrogen budget in the early PSN may not be 
simple: there are strong variations of the $^{14}$N/$^{15}$N isotopic ratio throughout the 
Solar System, which is indicative of the presence of more than one reservoir of 
volatile nitrogen \citep{Hily17}.
It is in fact possible that the N$_2$/CO ratio observed in 67P/C-G is not at all typical. 
Measurements for other comets indicate a range of 1-6\%, whilst recent observations of 
Comet C/2016 R2 (which has a long period of $\sim$20,550 years, originated in the Oort Cloud, 
has a semi-major axis of $\sim$1,500 AU and a perihelion distance of 2.6 AU) 
imply values of N$_2$:CO $\geq$ 6\%. The absence of any detection of NH$_2$ 
suggests that N$_2$ is the dominant reservoir for the nitrogen content of the ice 
\citep{Cochran18,Opetal19}. 
It is also worth noting that major trans-Neptunian objects (such as Triton or Pluto) are 
very 
N$_2$-rich.

\section{Discussion and Conclusions}
\label{sec:discussion}

Bearing in mind that there are several hypotheses that have been postulated to explain the 
high O$_2$ abundance in the coma of Comet 67P/C-G it is apparent that, despite the claims 
often made in the literature, the gas-grain chemistries of even simple molecules such as 
O$_2$ and H$_2$O are still highly uncertain and subject to debate.

The most successful previous studies have explained the relatively low abundance of O$_2$ 
in interstellar clouds as resulting from long static evolution, leading to the freeze-out 
of oxygen atoms and molecules and conversion to H$_2$O in normal circumstances \citep{Yildiz13}. 
The relatively high abundance of O$_2$ in comet 67P/C-G is then explained by conditions that 
yield an exceptionally low H:O ratio in the ices, resulting in the partial inhibition 
of the conversion of O and O$_2$ to H$_2$O \citep{TQ16}. 

Instead, we propose that sufficient O$_2$ may be produced by gas-phase 
reactions, and in a way that is robust to variations in certain free 
parameters and hence circumvents the necessity for these rather stringent
physical conditions.
In this scenario, the observed `O$_2$ excess' derives from a `fossil record', retained in 
the stratified ices, of the chemical and dynamical evolution of the molecular cloud from 
which the protosolar nebula formed.
Using a representative model of the time-dependent collapse of a primordial cloud 
into the pre-Solar
nebula, coupled to a gas-phase and gas-grain chemistry we find that 
a gas-phase primordial origin of cometary O$_2$ is both very viable and is fairly robust 
to the physical parameters (such as temperature and density). 

Our key findings for Comet 67P/C-G are:
\begin{itemize}
\item As a result of the non-linearities due to the competing timescales for gas-phase
chemistry, freeze-out and collapse, significant O$_2$ can be produced in transient peaks
in the collapse phase.
\item The O$_2$ is trapped in stratified ice mantles that retain a temporal record of 
the chemical evolution during the dynamical evolution of the cloud.
\item The outer ice-layers are O$_2$-poor, so that the low gas-phase and solid-state 
abundances that are inferred for molecular clouds are not incompatible with the existence of
a significant reservoir of hidden O$_2$.
\item Our results are robust to the choice of values for free parameters, but show
sensitivity to the assumed cosmic ray ionization rate and the rate of collapse.
The most favourable conditions for the formation of O$_2$ (and other species) at values 
that match observations are when the cosmic ionization rate is 0.1$\times$ the interstellar 
value (i.e. $\zeta\sim 10^{-18}$s$^{-1}$) and the collapse rate is ten times slower than 
free-fall.
\item Other abundance ratios, such as N$_2$:CO and CO:H$_2$O are also reasonably consistent with
observations.
\item The gas-phase origin provides a natural explanation for the relatively low observed
abundances of HO$_2$, H$_2$O$_2$ and O$_3$. 
\item Unless quite specific conditions prevail, the model significantly overproduces 
the N$_2$:H$_2$O ratio.
\item Several mechanisms have also been proposed in which the O$_2$
is created subsequent to sublimation from the cometary ice 
in the gas-phase coma chemistry. Whilst these have some drawbacks they cannot be excluded 
as possible sources of the O$_2$ excess.
\item Finally, it is quite possible that the chemical composition of the coma of Comet 67P/C-G is
anomalous and may not be representative of Jupiter-family comets.
\end{itemize}

On the assumptions that the O$_2$ abundances that are determined from the ROSINA data are 
representative of the composition of the cometary ice and are primordial in nature, 
then the results that we have obtained suggest that the observed enhancement of O$_2$ 
mainly originates from the phase in which the primordial cloud is collapsing to form the 
proto-solar nebula and is is not strongly sensitive to the physical conditions in that epoch.

% The large CO and CO$_2$ abundances relative to H$_2$O that we obtain (and which 
% were also reported by \citealp{TQ16}, when O$_2$ is relatively abundant) may be 
% at odds with the observation that the O$_2$ is more strongly-correlated to the H$_2$O 
% than it is to the CO, or CO$_2$......  

The relationship between the abundances in primordial molecular clouds and the pre-solar 
nebula is complex and not well-defined, as the abundance ratios will inevitably be 
modified by thermal processing in the early stages of the evolution of the Solar System.
Therefore, our results are somewhat compromised by the fact that pristine 
amorphous ices will almost certainly not survive intact into the protosolar nebula. 
Instead there will be a degree of thermal processing; heating, annealing and possibly 
re-freezing which will result in ices of varying compositions.
Thus, O$_2$ can be included, or trapped, in amorphous ices or clathrates - the latter being the more 
resilient to thermal processing. It seems like that some degree of sublimation, followed 
by re-freezing (possibly into clathrates) will occur but, in any case, the abundances of O$_2$ 
that we obtain must be regarded as upper limits for what will be retained in the ices in the
proto-Solar nebula.
This obviously leads to significant ambiguity in the determintion of the origin of 
the O$_2$ in Comet 67P/C-G.

The low observed abundance of N$_2$ relative to O$_2$ is surprising and is not predicted by the
primordial ice origin models. As there are no significant differences in 
the time-dependencies of the O$_2$ and the N$_2$ chemistries, and both species have very 
similar binding energies in the various ice phases, it is unlikely that this is an affect of 
compositional stratification in the ices. The fact that N$_2$ has been detected at significant 
levels in other comets suggests that the most likely cause is related to the specific thermal 
processing history of Comet 67P/C-G. 

\section*{Acknowledgements}

TGW acknowledges the financial support of the Science and Technology Facilities Council 
via a postgraduate studentship.

\bibliographystyle{mn2e}

%\bibliographystyle{mn2e_eprint}
%\bibliography{RollinsDiffuse2015}

\end{document}